\documentclass[aps,pra,amsmath,amssymb,showpacs,twocolumn]{revtex4-1}

\usepackage[T1]{fontenc}
\usepackage{amssymb,amsmath}
\usepackage{amsfonts}
\usepackage{graphicx}
\usepackage{color}
\usepackage{epstopdf}
\newcommand{\bJ}{{\bf J}}
\newcommand{\bn}{{\bf n}}

\newcommand{\tr}{{\rm tr}}

\def\ket#1{|#1\rangle}

\def\ketbra#1{|#1\rangle\langle#1|}

\begin{document}

\title{Absolutely classical spin states}
\author{F.~Bohnet-Waldraff$^{1,2}$, O.~Giraud$^{1}$, and D.~Braun$^{2}$}
\affiliation{
$^{1}$\mbox{LPTMS, CNRS, Univ. Paris-Sud, Universit\'e Paris-Saclay,
  91405 Orsay, France}\\
$^{2}$Institut f\"ur theoretische Physik, Universit\"at T\"{u}bingen,
72076 T\"ubingen, Germany 
}
\begin{abstract}
We introduce the concept of ``absolutely classical'' spin states, in
analogy to absolutely separable states of bi-partite quantum
systems. Absolutely classical states are states that remain classical
under any unitary transformation applied to them.  We
investigate the maximum ball of absolutely classical states 
centered on the fully mixed state that can be inscribed into the set
of classical states, and derive a lower bound for its radius as
function of the total spin quantum number. The result is compared to the 
case of absolutely separable states.
\end{abstract}

\pacs{03.65.Aa, 03.65.Ud, 03.67.-a}
\maketitle
\section{Introduction}

The rise of quantum information technology has led to the need to
classify and quantify the resources that ultimately enable a quantum
advantage in certain computational, communicational, or metrological
tasks. Most of the efforts have concentrated on classifying
entanglement.  Indeed, entanglement has been recognized to be
necessary for e.g.~computational speed-ups (at least for pure states)
\cite{Jozsa03}, 
quantum teleportation \cite{Bennett93}, super-dense coding
\cite{Harrow04}, and 
quantum data hiding \cite{DiVincenzo02}.  It  
can also be used for quantum key distribution \cite{Ekert91}, or for
achieving enhanced 
precision in certain metrological 
applications \cite{Giovannetti04}. 
  Recently, it has been
realized that other types of quantum mechanical correlations in the
form of ``quantum discord'' exist that do not require entanglement but
may still have useful applications \cite{OllZur01}.\\

Quantum entanglement necessarily requires at least bi-partite
systems.  However, even for a single system one can meaningfully ask
to what extent a particular quantum state shows genuine quantum
mechanical properties.  In quantum optics such questions were investigated
at least as early as the middle of last century. Quasi-probability
distributions were introduced that allow one to distinguish
``classical'' quantum states from states that show genuine
quantum effects such as enhanced quantum fluctuations of observables,
or quantum interference, including multi-photon
interference. An important role is played by coherent states of the
radiation field, in which the quantum fluctuations of the field
quadratures are minimal and evenly distributed over the canonical
coordinates.  Such states  come as close as quantum
mechanically possible to a point in classical phase space, and in
general, one can consider as classical states of the radiation field that can be expressed as a convex sum (i.e.~a classical mixture) of (projectors onto) coherent states
\cite{Mandel86,kim_nonclassicality_2005}.  Recently, these ideas were transferred to spin states, where SU(2) coherent states (introduced in \cite{Per72}) play the role of the most classical pure states, and a mixed spin state is considered ``classical'' if it can be written as a statistical mixture of SU(2) coherent states \cite{Giraud08}.  With this
classification, all states of a spin-1/2 are classical as they can be expressed as classical mixtures of pure states with minimal quantum fluctuations.  For a spin-1, there are genuinly nonclassical states, and necessary and sufficient conditions are known for classicality. These conditions can be used to explore analytically the geometry of quantum states \cite{QuantumnessSpin1} and provide a full analytical parametrization of the classical domain \cite{giraud_parametrization_2012}. For higher values of $j$, one can find sufficient conditions for non-classicality from the positivity of correlation functions of spin observables \cite{Giraud08,etapaper}. By definition, the classical states form a convex set, and one can define a ''quantumness'' measure of a state as the distance from this state to the convex set of classical states, in analogy to geometric measures of
entanglement \cite{Giraud10,Martin10}. Indeed, the two problems are related through the fact that spin-$j$ states can be also understood as states of $N=2j$ spins-1/2 fully symmetric under permutation of particles, so that quantumness of a spin-$j$ is equivalent to entanglement of $N=2j$ spins-1/2 in the fully symmetric sector of the Hilbert space of $N$ two-level systems \cite{etapaper}.  \\

Any measure of entanglement $E(\rho)$ is by definition invariant under local unitary
operations. But one can also ask for states for which $E(\rho)$ is
invariant under {\em any} 
unitary operation.  In particular, states $\rho$ such that $E(U\rho
U^\dagger)=0$ for all unitary operations $U$, called ``absolutely
separable'' states, have attracted substantial interest
\cite{kus_geometry_2001,separabilityWIKI,verstraete_maximally_2001,gurvits_largest_2002,zyczkowski_volume_1998,Arunachalam14,ganguly_witness_2014}. Absolutely
separable states have the property that no entanglement can be created
from them, no matter how strongly and how long the corresponding
particles interact. Conversely, for states which are not absolutely
separable there is at least in principle the possibility that some
entanglement be created from a common unitary evolution. The maximally
mixed state $\rho_0$, which is proportional to the identity matrix, is
obviously an absolutely separable state. As separable states form a closed set, there
is a ball around $\rho_0$ such that all states within that ball are
absolutely separable. Finding the largest radius of such a ball
provides a sufficient condition for  absolute separability; such a
question was addressed in \cite{gurvits_largest_2002}. 

In the present work, we ask an analogous question for quantumness:
what are the states of a spin-$j$ that remain classical no matter what
unitary evolution is applied to them? These states have the physical
interpretation that no quantumness can be created from  them in the
course of any unitary time evolution, generated  by an
arbitrary, even time-dependent hamiltonian. We correspondingly call
these states ``absolutely classical''.   
Alternatively,  states that
are not absolutely classical have the potential that in the course of
some unitary  evolution some quantumness may appear. 

The aim of the
paper is to provide a characterization of the set of absolutely
classical states in terms of a maximum distance from the maximally mixed spin-$j$ state, such that any state closer to the fully mixed state is guaranteed to be classical. This distance is the maximal radius that a ball of classical states around the maximally mixed spin-$j$ state can have. We provide a lower bound for this maximum radius based on an expansion of the Glauber-Sudarshan $P-$function into spherical harmonics, and calculate a numerical approximation by randomly sampling a large number of states and mixing them with the fully mixed state until their quantumness vanishes.  We start by defining the above concepts more precisely.

\section{Absolutely classical states}
\subsection{Classical spin states}
Pure classical spin states were defined in \cite{Giraud08} as SU(2)
coherent states.  This is motivated by the fact that these states have
minimal possible uncertainty of the angular momentum operator
$\bJ$. Moreover, when the spin undergoes a unitary time evolution
driven by a Hamiltonian linear in the components of $\bJ$,
corresponding for example to a precession in a magnetic field, this
minimal uncertainty property is conserved (similarly as what happens
for field coherent states, see e.g.~\cite{Arecchi72}). A spin-$j$
coherent state points in a well-defined direction $\bn$ that we
can parametrize with polar and azimuthal angles $\theta,\phi$ as
$\bn=(\sin\theta\cos\phi,\sin\theta\sin\phi,\cos\theta)$. In terms of
the usual $\ket{j,m}$ basis states (i.e.~eigenstates of $\bJ^2$ and
$J_z$ with eigenvalues $j(j+1)$ and $m$, respectively, with
$\hbar=1$), a spin-$j$ coherent state can be expanded as
\cite{Haake2000}  
\begin{equation}
\label{spin coherent}
\ket{\alpha} =\!\!\! \sum_{m=-j}^j \sqrt{\binom{2j}{j+m}} \left(\cos\frac{\theta}{2}\right)^{j+m}\left(\sin\frac{\theta}{2}e^{-i\phi}\right)^{j-m}\!\!\!\!\!\! \ket{j,m},
\end{equation}
with $\theta \in [0,\pi]$ and $\phi \in [0,2\pi[$. By a stereographic
projection $\alpha=e^{i\phi}\tan(\theta/2)$, we can alternatively
parametrize the 
spin-$j$ coherent states with a complex number $\alpha$. The $\ket{\alpha}$ form an
overcomplete basis and we have the identity $(1/4\pi)\int d\alpha \ketbra{\alpha}=I_{2j+1}$, where  $I_{2j+1}$ is the identity operator in the
$(2j+1)-$dimensional Hilbert space spanned by the $\ket{j,m}$ basis
states. Any density operator $\rho$ of a spin-$j$ state can be
expanded in terms of the $\ket{\alpha}$ in the form of a diagonal
representation,
\begin{equation}
  \label{eq:P}
  \rho=\int d\alpha P(\alpha)\ketbra{\alpha}\,,
\end{equation}
where $P(\alpha)$ is known as the (Glauber-Sudarshan) $P-$function
\cite{Agarwal81} (in general $P(\alpha)$ depends on both
$\alpha$ and $\alpha^*$, but it is customary to write $P(\alpha)$ for
short). \\ 

Classically mixing pure states should not increase their quantumness. This principle underlies the well-known definition of classicality in quantum optics
\cite{hillery_nonclassical_1987}.  In close correspondence, one can
therefore define mixed classical spin states as those states that can
be written as a classical mixture of spin-$j$ coherent states, i.e.~a
convex combination of projectors onto spin-$j$ coherent states. This
means that a general spin-$j$ state $\rho$ is classical iff there
exists a positive function $P(\alpha)$ with  which
$\rho$ can be written as in Eq.~\eqref{eq:P}  \cite{Giraud08}. Note
that the $P$-function is not unique: indeed, when 
expanded over spherical harmonics $Y_{KQ}(\theta,\phi)$, only
components with $K\leqslant 2j$ 
 play a role in the integral \eqref{eq:P}, so that arbitrary spherical
 harmonics with $K\geqslant 2j+1$ 
can be added to $P(\alpha)$ without changing $\rho$ (see an example in
\cite{Giraud08}). Classical spin-$j$ states are hence those states for which at
                   least one $P-$function is positive. 
Classical spin-$j$ states form a convex set by definition. 
Deciding whether a spin-$j$ state is classical or not then becomes a
problem of convex optimization (see below). 
Note that the Wigner 
function of a spin-$j$ coherent state is not everywhere  positive in
general,  not even for a spin-1/2, see
\cite{AgarwalQOpticsbook}. 
This is different from the harmonic
oscillator, where positivity of the $P$-function implies positivity of
$W$.\\

\subsection{Absolutely classical spin states}
Let $\mathcal{H}_m$ be a Hilbert
space of dimension $d_m$, and $\mathcal{B}(\mathcal{H}_m)$ the space
of
bounded linear operators on $\mathcal{H}_m$.  Consider a bipartite
physical system with Hilbert space
$\mathcal{H}=\mathcal{H}_m\otimes\mathcal{H}_n$. 
In \cite{kus_geometry_2001} the absolute separability problem was
introduced:
What are the states $\rho\in\mathcal{B}(\mathcal{H})$  such that $U\rho  U^\dagger$
is separable for all unitary matrices  $U\in
\mathcal{B}(\mathcal{H})$? The problem can also be
understood as ``separability  from spectrum''-problem
\cite{separabilityWIKI}: Since all   $U\rho U^\dagger$
have the same spectrum of eigenvalues  as $\rho$, it is natural to try
to characterize the set of absolutely separable states by conditions
on the spectrum. For $n=m=2$, a necessary  and sufficient condition is
known in terms of a single inequality for  the eigenvalues
\cite{verstraete_maximally_2001}: if
$\lambda_1\geqslant \lambda_2\geqslant \lambda_3\geqslant \lambda_4$ are the eigenvalues
of $\rho$, then it is absolutely separable if and only if
$[(\lambda_1-\lambda_3)^2+(\lambda_2-\lambda_4)^2]^{1/2}\leqslant \lambda_2+\lambda_4$. 
Absolute separability is evidently a stronger condition than
separability. For instance a coherent state of two spins-1/2 is a
separable state, but it can become entangled under a general unitary
transformation $U\in \mathcal{B}(\mathcal{H})$. More generally, no
pure two-qubit state satisfies the above inequality, hence any
two-qubit  absolutely separable state is mixed.
The general problem is still  open. \\

In \cite{gurvits_largest_2002} an
important step was made by  finding the largest ball of separable
states (in terms of any $p$-norm, $0\leqslant p\leqslant \infty$) centered at
the   maximally mixed state $\rho_0=I_m\otimes I_n/d$ with $d=mn$. In
the Frobenius  norm ($p=2$) its radius is  given by
$r_d=1/\sqrt{d(d-1)}$, i.e.~all $\rho$ with $||\rho-\rho_0||\leqslant r_d$
are separable, and $r_d$ is the largest such  constant. In terms of
purity this means that $\rho$ is separable if  the purity $\tr\rho^2$
is less than or equal to $1/(d-1)$, as was already  conjectured in
\cite{zyczkowski_volume_1998}.   Although all states within this ball
are absolutely separable, there are also  absolutely separable states
outside this ball \cite{Arunachalam14}. This can be clearly seen in
the case $n=m=2$: it is easy to find examples of states $\rho$ whose
distance to $\rho_0$ in the Frobenius norm satisfies 
$[\sum_i(\lambda_i-1/4)^2]^{1/2}>r_d=1/\sqrt{12}$, while the absolute
separability condition
$[(\lambda_1-\lambda_3)^2+(\lambda_2-\lambda_4)^2]^{1/2}\leqslant \lambda_2+\lambda_4$
is satisfied (for instance $\lambda_1=\lambda_2=13/32$ and
$\lambda_3=\lambda_4=3/32$). Witnesses for states that are not
absolutely separable  were introduced in \cite{ganguly_witness_2014}.  \\

Here we ask a corresponding question for classicality: 
what are the   spin-$j$ states $\rho\in \mathcal{B}(\mathcal{H}_{2j+1})$ such that
$U\rho  U^\dagger$ is classical for all unitary matrices  $U\in
\mathcal{B}(\mathcal{H}_{2j+1})$? The states that fulfill this criterion  will be
called ``absolutely classical''. They are such that no unitary
spin-$j$ operator can create quantumness, or equivalently, 
 entanglement among the underlying $N=2j$ spins-1/2. We proceed
 similarly to the approach of \cite{gurvits_largest_2002}, i.e.~we
 establish a lower bound on the maximal radius $r_{\rm max}(j)$ of the
 ball around the maximally mixed state
 $\rho_0=I_{2j+1}/(2j+1)$, in which  any state is classical.

\subsection{Analytical lower bound for $r_{\rm max}(j)$}
Let $\rho$ be an arbitrary density matrix of a spin-$j$ state. This state can always be written as 
\begin{equation}
\label{statedef}
\rho(r)=\rho_0+r\tilde{\rho},
\end{equation} 
where $\rho_0={ I}_{2j+1}/(2j+1)$, 
and
$\tilde{\rho}=(\rho-\rho_0)/||\rho-\rho_0||$  is traceless and
normalized so that the (Hilbert-Schmidt or Frobenius) norm of
$\tilde{\rho}$ is $||\tilde{\rho}||^2=\tr\tilde{\rho}^2=1$ without restriction of
generality. This fixes the scale  for the real positive parameter
$r$. Therefore the state $\rho(r)$ is at the distance $r$ from the
maximally mixed state. The $P$-function of $\rho(r)$,  defined through
the coherent state representation   
\begin{equation} \label{prep} 
\rho(r)=\int d\alpha P(r,\alpha)|\alpha\rangle\langle\alpha|\,,
\end{equation}
can be written 
\begin{equation} \label{p}
P(r,\alpha)=\frac{1}{4\pi}+r \tilde{P}(\alpha)\,,
\end{equation}
where $1/4\pi$ is the $P$-function of $\rho_0$. 
In order to show that for a given $r$ and arbitrary direction
$\tilde{\rho}$ a positive $P$-function can be found, it is enough to
consider traceless parts that 
can be expanded as
\begin{equation} \label{tp}
\tilde{P}(\alpha)=\sum_{K=1}^{2j}\sum_{Q=-K}^K \tilde{P}_{KQ}Y_{KQ}(\alpha)\,,
\end{equation}
where the $Y_{KQ}$ are spherical harmonics and $\tilde{P}_{KQ}\in
\mathbb{C}$. 
Note that more generally 
$\tilde{P}(\alpha)$ can contain spherical harmonics with arbitrarily
large $K$, but any $\rho(r)$ 
can be represented by a $P$-function that contains values of $K$ only
up to $2j$. Indeed, a given quantum state fixes the components in
$P(\alpha)$ up to
$K=2j$ uniquely (see below), whereas the higher ones are arbitrary.
Hence we can set them to zero and look for the largest $r$ that still
guarantees for all $\tilde{\rho}$ a positive $P(\alpha)$ of the form \eqref{tp}.
We can expand $\tilde{\rho}$ in terms of
the irreducible tensor operators $T_{KQ}$ as 
\begin{equation} \label{irr}
\tilde{\rho}=\sum_{K=1}^{2j}\sum_{Q=-K}^K\tilde{\rho}_{KQ}T_{KQ}\,.
\end{equation}
Completely analogously, we can also
expand $\rho(r)$ and $P(r,\alpha)$ in terms of $T_{KQ}$ and
$Y_{KQ}(\alpha)$, respectively:
\begin{eqnarray} \label{irr}
\rho(r)&=&\sum_{K=0}^{2j}\sum_{Q=-K}^K\rho_{KQ}(r)T_{KQ}\,,\\
P(r,\alpha)&=&\sum_{K=0}^{2j}\sum_{Q=-K}^K P_{KQ}(r)Y_{KQ}(\alpha)\,.
\end{eqnarray}
One then immediately finds $P_{KQ}(r)=r \tilde{P}_{KQ}$ and
$\rho_{KQ}(r)=r \tilde{\rho}_{KQ}$ for all integer $K\geqslant 1$ and $-K\leqslant 
Q\leqslant K$. 
Since $\rho(r)$ is a valid density matrix, the $P_{KQ}(r)$ are
related to the $\rho_{KQ}(r)$ by a simple factor
\cite{agarwalTensorRep}, 
\begin{align} \label{bP}
P_{KQ}(r)=f_{KQ}\,\rho_{KQ}(r)\,\,\, 
 \forall\,\,\, K,Q\,,\\
f_{KQ}=(-1)^{K-Q}\frac{\sqrt{(2 j-K)! (2 j+K+1)!}}{2 \sqrt{\pi } (2 j)!}\,,
\end{align}
and hence also 
\begin{equation} \label{bP}
\tilde{P}_{KQ}=f_{KQ}\,\tilde{\rho}_{KQ}
\end{equation}
$\forall K\geqslant 1,\,-K\leqslant Q\leqslant K$.
Cauchy-Schwarz inequality applied to \eqref{tp} then yields
\begin{multline} \label{Phatbound}
|\tilde{P}(\alpha)|\leqslant \left(\sum_{K=1}^{2j}\sum_{Q=-K}^K
  |\tilde{\rho}_{KQ}|^2\right)^{1/2} \\
\times\left(\sum_{K=1}^{2j}\sum_{Q=-K}^K
  | f_{KQ}Y_{KQ}(\alpha)|^2\right)^{1/2}\,.
\end{multline}
The normalization of $\tilde{\rho}$ implies 
\begin{align} 
\begin{split}
\label{tr2}
\sum_{K=1}^{2j}\sum_{Q=-K}^K|\tilde{\rho}_{KQ}|^2&=\sum_{K,K=1'}^{2j}\sum_{Q,Q'=-K}^K\tilde{\rho}_{KQ}\tilde{\rho}^*_{K'Q'}\tr
T_{KQ} T_{K'Q'}^\dagger \\
&=\tr\tilde{\rho}^2=1\,,
\end{split}
\end{align}
where we have used the orthogonality of the irreducible tensor operators.
By noting that $|f_{KQ}|$ is independent of $Q$ and using the identity
\begin{equation}
\sum _{Q=-K }^{K }|Y_{K Q}(\theta ,\varphi )|^2={\frac {2K +1}{4\pi }},
\end{equation}
we get from \eqref{Phatbound} that
$|\tilde{P}(\alpha)|\leqslant \tilde{P}^{(j)}_\text{max}$, with
\begin{align}
 \tilde{P}^{(j)}_\text{max}=\left\{\frac{2j+1}{8\pi^2}\left[(4j+1){4j
    \choose 2j}-(j+1)\right]\right\}^{1/2}\,.\label{minP}
\end{align}
This implies a lower bound 
\text{$\tilde{P}(\alpha)\geqslant -\tilde{P}_\text{max}^{(j)}$}, and hence 
\begin{equation}
P(r,\alpha)= \frac{1}{4\pi} +r \tilde{P}(\alpha) \geqslant \frac{1}{4\pi} -r
\tilde{P}_{\rm max}^{(j)}. 
\end{equation}
If the right-hand side is non-negative, so is the  left-hand side. Thus if
\begin{align} \nonumber
  r\leqslant \frac{1}{4 \pi \tilde{P}_{\rm max}^{(j)}}&=\left\{(4 j+2)
    \left[(4 j+1) \binom{4 j}{2 j}-(j+1)\right]\right\}^{-1/2}\\ 
 & \equiv \hat{r}_{\rm
    max}(j)\label{eq:bound} \,,
\end{align}
 in the state \eqref{statedef}, then the $P-$function, given by
 $P(r,\alpha)$ in Eq.~\eqref{p} is positive.  
Hence, $\rho(r)$  is 
classical for  $r\leqslant \hat{r}_{\rm
    max}(j)$.  Since $\rho(r)=\rho$ for $r=||\rho-\rho_0||$, we have
  proved that
\begin{equation} 
\label{lowerbound}
||\rho-\rho_0||\leqslant \hat{r}_{\rm max}(j)\Rightarrow \rho\in {\cal C}\,, 
\end{equation}
where ${\cal C}$ is the set of classical states. The distance
$||\rho-\rho_0||$  is invariant under conjugation by an arbitrary
unitary matrix $U\in\mathcal{B}(\mathcal{H}_{2j+1})$. Hence, if $\rho$
satisfies the inequality in \eqref{lowerbound}, all states 
$U\rho U^\dagger$ verify $||U\rho U^\dagger -\rho_0||\leqslant 
\hat{r}_{\max}(j)$ and are thus classical. Therefore
$\hat{r}_{\rm max}(j)$ is a lower bound for the ball size 
$r_\text{max}(j)$.  \\

The Cauchy-Schwarz inequality \eqref{Phatbound} can be saturated
for any given $\alpha$ by choosing
$\tilde{\rho}_{KQ}=Af_{KQ}Y_{KQ}(\alpha)$ where $A$ is a
proportionality constant such that $\tr\tilde{\rho}^2=1$. However, due
to the restriction of the $P$-function to \eqref{tp}, with components
$K\leqslant 2j$ only, we do not exhaust all possible $P$-functions. Hence, it
may be possible  to increase the lower bound of $\tilde{P}(\alpha)$ in
\eqref{minP} by adding components $Y_{KQ}$ with $K>2j$. 

\subsection{Numerical result for $r_{\rm max}(j)$}
To test the lower bound \eqref{eq:bound}, we search for non-classical
states that are as close as possible to the maximally mixed state, since
each of these states gives an upper bound on the true ball size
$r_{\rm max}(j)$.   

To do this, we generate random mixed states $\rho$ from the
Hilbert-Schmidt ensemble of matrices 
$\rho=AA^{\dagger}/\tr(AA^{\dagger})$, with $A$ a complex matrix with
independent Gaussian entries (see \cite{zyczkowski_generating_2011} for details). With these states, we construct families of states 
\begin{equation}
\label{StartState}
\rho_k=(1-k) \rho_0  + k \rho,
\end{equation}
as function of a parameter $k \in [0,1]$, that interpolate between the
maximally mixed state $\rho_0$ and the state $\rho$. The task is to
find the largest value $k_{\max}$ of $k$, under the condition that $\rho_k$ is
classical. This can be rewritten as 
\begin{equation}
\label{kMaximaisationProblem}
\max_k k \quad \mbox{s.t.}\quad \rho_k=\int d\alpha P(\alpha)|\alpha\rangle\langle\alpha|, \, P(\alpha)\geqslant 0.
\end{equation}

This problem can be formulated in the form of a linear programming problem, of the form
\begin{align}
\label{LinearAlgo}
\max_x c^Tx \quad \mbox{s.t.} \quad Ax=b, \, x\geqslant 0,
\end{align}
where $x$ is the vector of variables, $c,b$ are real given vectors and $A$ is a real given matrix. 
These types of optimizations can be solved very efficiently e.g. with
an interior-point method \cite{VanBoy96}. Another great property is the
existence of a dual problem. If the optimal value of the dual problem
coincides with the optimal value of the original problem
\eqref{LinearAlgo}, i.e.~if there is no duality gap the solution is proven to be optimal. We will now explain how 
to reformulate the problem \eqref{kMaximaisationProblem} in the form
\eqref{LinearAlgo}.  \\

Due to Carath\'eodory's theorem, a
positive $P-$function for finite $j$ can
always be written as a convex sum of delta functions, so any classical
state has the form 
$\sum_{i=1}^N w_i \ketbra{\alpha_i}$
with $w_i \geqslant 0 $ and $N\leqslant (2j+1)^2$ (where the number of states
needed is reduced by one due to normalization of the
state). 
With this form, $\rho_k$ is
classical iff there exist $w_i\geqslant 0$ with $\sum_i w_i=1$
and coherent states $\ket{\alpha_i}$ such that   
\begin{equation}
\sum_{i=1}^N w_i\ketbra{\alpha_i}=\rho_k\,,
\end{equation}
which can be rewritten as
\begin{equation}
\label{linalgexplict}
\sum_{i=1}^N w_i \ketbra{\alpha_i} +k  \left(\rho_0-\rho\right)=\rho_0\,. 
\end{equation}
This equation can be written as $Ax=b$ as in \eqref{LinearAlgo}, where
the vector of variables is given by
$x=(\{w_i\}_{i=1,\ldots,N},k)$. 
  The vector $b$ is fixed
by the maximally mixed state, and the matrix $A$ is constructed from
the real and imaginary entries of the left-hand side of
Eq.\eqref{linalgexplict}.   Then with the choice $c=(0,\ldots,0,1)$ in 
\eqref{LinearAlgo}, the problem \eqref{kMaximaisationProblem} is in
the form of a linear optimization problem.

However, since the
$\alpha_i$ in \eqref{linalgexplict} are unknown, we generate a large list of
uniformly distributed coherent states, of the order of $10^6$ many, so that
it should be possible to construct almost all classical states by
varying the weights $w_i$. This assumption can be tested by 
repeating the linear optimization  with a new set of random angles and
also with an increased number of them. These tests showed that for
$j\leqslant 21/2$, increaseing the number of random angles beyond $10^6$
does not visibly change the results. 

We applied this procedure to a list of $n\sim 3000$ 
different states $\rho$ in  
 \eqref{StartState} for system sizes of up to
$j=21/2$. The states that maximize $k$ are found 
at distances 
 \begin{equation}
 \label{GeneratingRList}
r_l=\left|\left|\rho_{\tilde{k}_{\max}} - \rho_0 
  \right|\right|=\left|\left|\rho-\rho_0
  \right|\right| \tilde{k}_{\max}, \,\,\,l=1,2,\ldots, n 
 \end{equation}
from the fully mixed spin-$j$ state, where $\tilde{k}_{\max}$ is
the numerical result of the optimization problem
\eqref{LinearAlgo},\eqref{linalgexplict}.      
Numerically, not all directions $\rho$ can
be sampled, and only a finite number of coherent states can be
considered.   On the one hand, the fact that we can sample  only  a
finite number of coherent states entails that the numerically found
$\tilde{k}_{\max}$ for a given $\rho$ is a lower
bound of the corresponding exact $k_{\max}$.  On the other hand, even if one
started with all coherent states,  
as in the decomposition  
\eqref{kMaximaisationProblem}, one would achieve the exact values
$\tilde{k}_{\max}=k_{\max}$, but each $r_l$, and hence $\tilde{r}_{\rm
  max}(j)\equiv \min_{1\leqslant l\leqslant 
  n}r_l$, would still give only an upper bound on the true   
radius of the ball $r_{\max}(j)$. 
Therefore, 
$\tilde{r}_{\max}$  is simply a numerical approximation of $r_{\max}$,
but {\em a priori} neither a strict upper nor lower bound.  

It is worth mentioning that the entangled states
closest to the maximally mixed state are not on a straight 
line with the queen
of quantum state, i.e. the state with maximum quantumness for given
$j$ \cite{Giraud10}, except in the $j=1$ case.

\begin{figure}[h!]
\begin{center}
\includegraphics[width=0.48\textwidth]{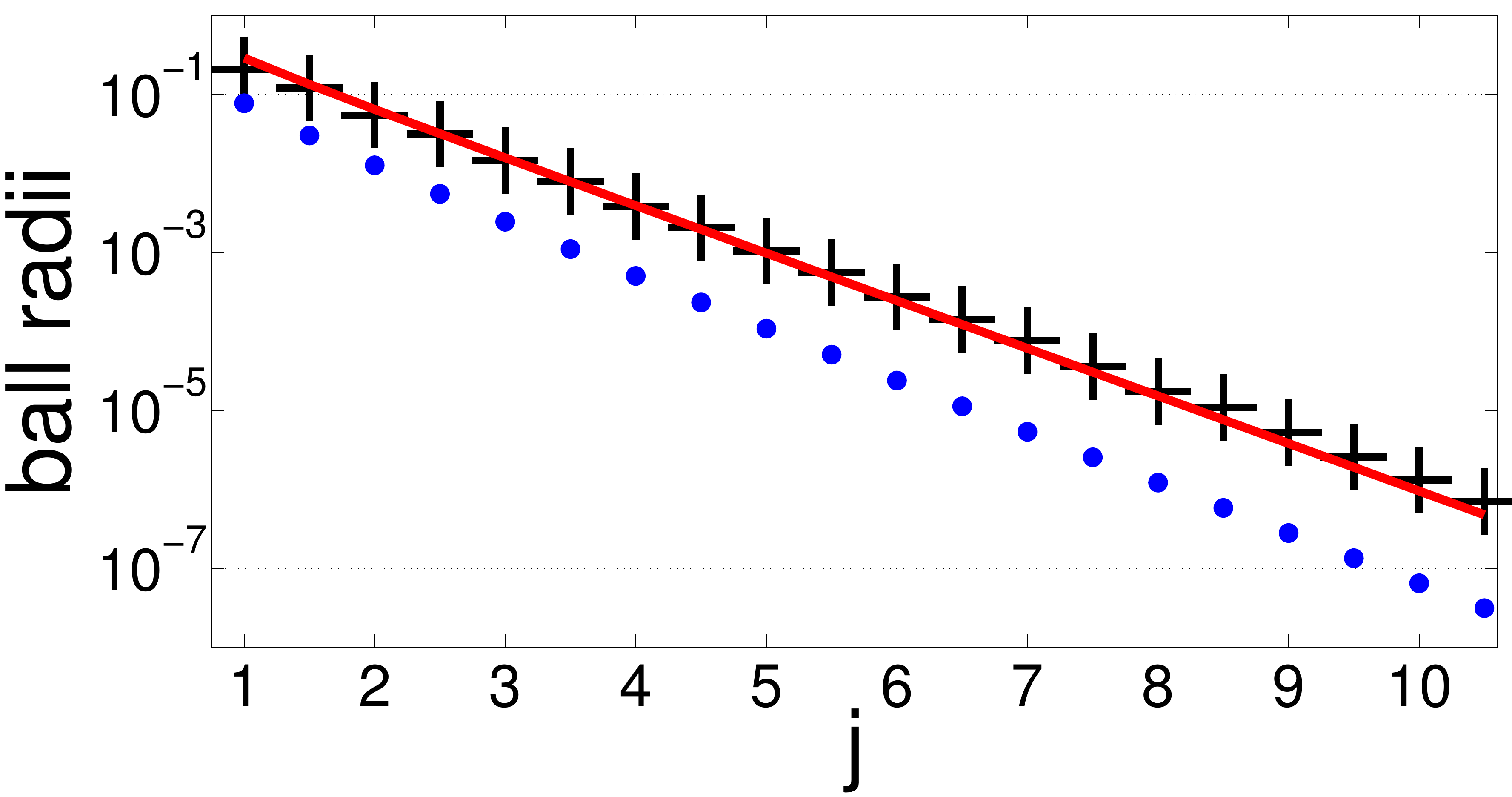}
\end{center}
\caption{{\em (Color online).} Maximal radius $r_{\max}(j)$ of a ball of
  classical states centered at the fully mixed state as function of $j$. 
 Blue dots: The value of the lower
  bound $\equiv \hat{r}_{\rm
    max}(j)$, Eq.~\eqref{eq:bound}. 
Black crosses: smallest numerically found
  distance from the 
  maximally mixed state to a non-classical state. Red line: maximal
  ball size $1/(2^j \sqrt{4^j-1})$ for arbitrary (not necessarily symmetric)
  separable states 
  \cite{gurvits_largest_2002}. 
  This function gives an excellent approximation
  of the numerically found maximal ball size $\tilde{r}_{\rm max}(j)$
  of classical spin-$j$ states, but slightly overestimates it for 
  small $j$.} 
\label{fig:lowerboundandData}
\end{figure}

\section{Discussion}
In Fig.~\ref{fig:lowerboundandData} we compare the numerically found
$\tilde{r}_{\rm max}(j)$ and the analytical
lower bound $\hat{r}_{\rm max}(j)$ from Eq.~\eqref{eq:bound} with the
radius of the ball of absolutely separable states,
$r_d=1/(2^j\sqrt{4^j-1})$ with $d=2^{2j}$ \cite{gurvits_largest_2002}.
The lower bound $\hat{r}_{\rm max}(j)$ decays exponentially with $j$. It
is still substantially below the 
numerically found 
$\tilde{r}_{\rm max}(j)$, which can be
considered close to the exact value $r_{\rm max}(j)$. Also $\tilde{r}_{\rm
  max}(j)$ decays exponentially with $j$, and the ratio between
$\tilde{r}_{\rm max}(j)$ and $\hat{r}_{\rm max}(j)$ increases only slowly
with increasing $j$ over the whole examined range $1\leqslant j\leqslant 10.5$. The
function $1/(2^j\sqrt{4^j-1})$ agrees with $\tilde{r}_{\rm max}(j)$ remarkably well
over the whole range of $\rho$.  However, it is not to be expected
that $1/(2^j\sqrt{4^j-1})$ is the correct result for $r_{\rm
  max}(j)$ for at least two reasons:  ({\em i.}) the fully mixed
state in the fully 
symmetric sector of Hilbert space ${ I}_{2j+1}/(2j+1)$ 
 (under exchange of
qubits)  is not identical to the fully mixed state in the full Hilbert
space $\mathcal{H}$, ${ I}_{2^{2j}}/2^{2j}$, of $N=2j$ spins-1/2. 
Hence, the balls of absolutely
separable states and absolutely classical states are not centered at
the same point. For example for two spins-1/2, we have a
fully symmetric subspace of $\mathcal{H}$ of dimension 3 (the triplet
sector) with the identity matrix ${
  I}_3\equiv\sum_{m=-1}^1\ketbra{1,m}$, 
whereas the identity in the
full $\mathcal{H}$ also contains a projector onto the singlet state
$\ketbra{j=0,m=0}$, and has hence to be normalized 
differently as well, 
${
  I}_4\equiv\sum_{j=0,1}\sum_{m=-j}^j\ketbra{j,m}$. And ({\em ii.)}, 
when minimizing the distance to non-classical states, the relevant set
of states is larger without the restriction to symmetric states. From
the latter argument one would expect that $1/(2^j\sqrt{4^j-1})$
underestimated $r_{\rm max}(j)$, if it were evaluated centered on the
same identity.  This appears to be correct for large
values of $j$ (starting at about $j\geqslant 4$), but could there also be due
to the numerical uncertainty of the very small value of $r_{\rm
  max}(j)$. 
For small values of $j$, we have rather $\tilde{r}_{\rm
  max}(j)<1/(2^j\sqrt{4^j-1})$.  The case $j=1$ is particularly instructive,
as there we have a full analytical characterization of the set of
classical states \cite{giraud_parametrization_2012}.  The numerically found value $\tilde{r}_{\rm
  max}(j)\simeq 0.2052$  agrees well with the analytical one
$1/( 2\sqrt{6})\simeq 0.2041$ whereas $1/(2^j\sqrt{4^j-1})=1/(2\sqrt{3})\simeq 0.288$. 
Nevertheless, altogether we see that the closest non-classical symmetric
state of a spin-j is about as close to the fully mixed state in the symmetric
sector as the closest entangled state without any symmetry
restrictions to the fully mixed state in the full $2^{2j}$ dimensional
Hilbert space of $N=2j$ spins-1/2.

{\bf Acknowledgments:} DB
thanks OG, the LPTMS, and the Universit\'e Paris Saclay for
hospitality. We thank Peter Braun for useful discussions, and the Deutsch-Franz\"osische 
Hochschule (Universit\'e franco-allemande) for support, grant
number CT-45-14-II/2015. Ce travail a b\'en\'efici\'e
d'une aide Investissements d'Avenir du LabEx PALM
(ANR-10-LABX-0039-PALM).



\begin{thebibliography}{99}
\bibitem{Jozsa03} R.~Jozsa and N.~Linden, Proc. R. Soc. Lond. A {\bf 459}, 2011 (2003).

\bibitem{Bennett93} C.~H.~Bennett, G.~Brassard, C.~Crepeau, R.~Jozsa, A.~Peres and W.~K.~Wootters, Phys. Rev. Lett. {\bf 70}, 1895 (1993).
  

\bibitem{Harrow04} A.~Harrow, P.~Hayden, and D.~Leung, Phys. Rev. Lett. {\bf 92}, 187901 (2004).
 

\bibitem{DiVincenzo02} D.~P.~DiVincenzo, D.~W.~Leung, and B.~M.~Terhal, IEEE Trans. Inf. Theory {\bf 48}, 580 (2002). 

\bibitem{Ekert91} A.~K.~Ekert, Phys. Rev. Lett. {\bf 67}, 661 (1991).
 

\bibitem{Giovannetti04} V.~Giovannetti, S.~Lloyd, and L.~Maccone, Science {\bf 306}, 1330 (2004).
 

\bibitem{OllZur01} H.~Ollivier and W.~H.~Zurek, Phys.~Rev.~Lett.~{\bf 88}, 017901 (2001).

\bibitem{Mandel86} L.~Mandel, Phys. Scr. {\bf T12}, 34 (1986).
  

\bibitem{kim_nonclassicality_2005} M.~S.~Kim, E.~Park, P.~L.~Knight, and H.~Jeong, Phys. Rev. A {\bf 71}, 43805 (2005).
 

\bibitem{Per72} A.~M.~Perelomov, Commun.~Math.~Phys.~{\bf 26}, 222--236 (1972).

\bibitem{Giraud08} O.~Giraud, P.~Braun, and D.~Braun, Phys. Rev. A {\bf 78}, 42112 (2008).
 

\bibitem{QuantumnessSpin1} F.~Bohnet-Waldraff, D.~Braun, and O.~Giraud, Phys. Rev. A {\bf 93}, 012104 (2016).
 

\bibitem{giraud_parametrization_2012} O.~Giraud, P.~Braun, and D.~Braun, Phys. Rev. A {\bf 85}, 032101 (2012).
 

\bibitem{etapaper} F.~Bohnet-Waldraff, D.~Braun, and O.~Giraud, Phys. Rev. A {\bf 94}, 42343 (2016).
 

\bibitem{Giraud10} O.~Giraud, P.~Braun, and D.~Braun, New J.~Phys.~{\bf 12}, 063005 (2010).
 

\bibitem{Martin10} J.~Martin, O.~Giraud, P.~A.~Braun, D.~Braun, and T.~Bastin, Phys. Rev. A {\bf 81}, 62347 (2010).
 

\bibitem{kus_geometry_2001} M.~Ku\'s and K.~\.{Z}yczkowski, Phys. Rev. A {\bf 63}, 32307 (2001).
 

\bibitem{separabilityWIKI} "Separability from spectrum" - OpenQIProblemsWiki http://qig.itp.uni-hannover.de/qiproblems/15.
 

\bibitem{verstraete_maximally_2001} F.~Verstraete, K.~Audenaert, and B.~De~Moor, Phys. Rev. A {\bf 64}, 12316 (2001). 

\bibitem{gurvits_largest_2002} L.~Gurvits and H.~Barnum, Phys. Rev. A {\bf 66}, 62311 (2002).

\bibitem{zyczkowski_volume_1998} K.~\.{Z}yczkowski, P.~Horodecki, A.~Sanpera, and M.~Lewenstein, Phys. Rev. A {\bf 58}, 883 (1998).

\bibitem{Arunachalam14} S.~Arunachalam, N.~Johnston, and V.~Russo, Quantum Inform. Comput. {\bf 15}, 0694, (2015)

\bibitem{ganguly_witness_2014} N.~Ganguly, J.~Chatterjee, and A.~S.~Majumdar, Phys. Rev. A {\bf 89}, 52304 (2014).

\bibitem{Arecchi72} F.~T.~Arecchi, E.~Courtens, R.~Gilmore, and H.~Thomas, Phys.~Rev.~A {\bf 6}, 2211 (1972).

\bibitem{Haake2000} F.~Haake, \textit{Quantum Signatures of Chaos}, (Springer, Berlin, Heidelberg, 2010).

\bibitem{Agarwal81} G.~S.~Agarwal, Phys. Rev. A {\bf 24}, 2889 (1981).

\bibitem{hillery_nonclassical_1987} M.~Hillery, Phys.~Rev.~A {\bf 35}, 725 (1987).

\bibitem{AgarwalQOpticsbook} G.~S.~Agarwal, \textit{Quantum Optics}, (Cambridge University Press, Cambridge, 2012).

\bibitem{agarwalTensorRep} G.~S.~Agarwal, Phys. Rev. A {\bf 47}, 4608 (1993).
 

\bibitem{zyczkowski_generating_2011} K.~\.{Z}yczkowski, K.~A.~Penson, I.~Nechita, and B.~Collins, J.~Math.~Phys.~{\bf 52}, 062201 (2011).
 

\bibitem{VanBoy96} L. Vandenberghe and S. Boyd, SIAM Rev. {\bf 38}, 49 (1996).
 

\end{thebibliography}
\end{document}